\documentclass[11pt]{article}

\usepackage[T1]{fontenc}
\usepackage[utf8]{inputenc}
\usepackage[a4paper,margin=1in]{geometry}
\usepackage{amsmath,amssymb}
\usepackage[hidelinks]{hyperref}

\newcommand{\BY}[1]{#1,}
\newcommand{\TITLE}[1]{\emph{#1}}
\makeatletter
\newcommand{\IN}[4]{#1%
  \if\relax\detokenize{#2}\relax\else\ \textbf{#2}\fi%
  \if\relax\detokenize{#3}\relax\else\ (#3)\fi%
  \if\relax\detokenize{#4}\relax\else, #4\fi}
\makeatother

\title{Advances in quantum learning theory with bosonic systems}
\author{Francesco Anna Mele\thanks{\href{mailto:francesco.mele@sns.it}{\texttt{francesco.mele@sns.it}}}\\
\small Scuola Normale Superiore, Pisa, Italy}
\date{}

\begin{document}

\maketitle

\begin{abstract}
This paper reviews recent advances in quantum learning theory for continuous-variable (CV) systems. Quantum learning theory investigates how to extract classical information from quantum systems as efficiently as possible. CV systems are ubiquitous in nature and in quantum technologies, as they describe bosonic and quantum-optical systems. While quantum learning theory for finite-dimensional systems has been extensively studied, the corresponding theory for CV systems has only recently begun to develop; here we provide a concise review. We focus on the following questions: what is the minimum number of copies (the sample complexity) required to learn a non-Gaussian state, possibly under energy constraints? What is the sample complexity for learning Gaussian states? How does the performance of CV state learning depend on non-Gaussianity? How can one test whether a state is Gaussian or far from the set of Gaussian states? And how can Gaussian processes be learned efficiently? Central to these topics, we also review several bounds on the trace distance between CV states in terms of their covariance matrices, which may be of independent interest. Overall, this work summarises selected developments in tomography of CV systems and highlights a selection of open problems.
\end{abstract}

\section{Introduction}
Quantum learning theory~\cite{ref:AA24} is a relatively new area of quantum information that studies how to extract classical information from quantum systems as efficiently as possible. A central problem in quantum learning theory is \emph{quantum state tomography}~\cite{ref:AA24,ref:OW16,ref:HHJ17,ref:PSTW25}, i.e., the task of learning an unknown quantum state.

More precisely, tomography can be stated as follows: you are given $N$ copies of an unknown quantum state $\rho$, promised to lie in a set $\mathcal{S}$. Fix an error probability $\delta\in(0,1)$ and a trace-distance~\cite{ref:Helstrom76,ref:Holevo76Steklov} accuracy parameter $\varepsilon$. The goal is to take as input $\rho^{\otimes N}$ into a quantum algorithm that outputs a classical description of an estimator $\tilde{\rho}$ such that $\Pr\!\left[\frac12\|\rho-\tilde{\rho}\|_1 \le \varepsilon\right] \ge 1-\delta$. Here the probability is over the internal randomness of the procedure (equivalently, over the random output $\tilde{\rho}$), and $\frac12\|\rho-\tilde{\rho}\|_1$ denotes the trace distance~\cite{ref:Helstrom76,ref:Holevo76Steklov} between the true state $\rho$ and the estimate $\tilde{\rho}$. For fixed $\varepsilon,\delta\in[0,1]$ and prior set $\mathcal{S}$, the minimum number of copies $N$ that suffices to solve the problem of tomography is called the \emph{sample complexity}, and it is the main object of study in quantum learning theory. For example, when $\mathcal{S}$ is the set of all mixed states of a qudit of dimension $D$, the sample complexity is $N=\Theta\!\left(\frac{D^2+\log(1/\delta)}{\varepsilon^2}\right)$~\cite{ref:OW16,ref:HHJ17,ref:PSTW25}. If instead $\mathcal{S}$ is the subset of all \emph{pure} states of a qudit of dimension $D$, the sample complexity reduces to $N=\Theta\!\left(\frac{D+\log(1/\delta)}{\varepsilon^2}\right)$~\cite{ref:OW16,ref:HHJ17,ref:PSTW25,ref:GKKT20}.

Other natural questions in quantum learning theory include how to estimate expectation values of many observables as efficiently as possible---a task known as shadow tomography~\cite{ref:Aaronson18,ref:HKP20}. One can also ask how to learn unitaries~\cite{ref:HKOT23}---or, more generally, quantum channels in diamond distance~\cite{ref:MeleBittel26, ref:CYZ26}. While the topic of quantum learning theory has developed extensively over the last decade for finite-dimensional systems, only recently has quantum learning theory for continuous-variable (CV) systems~\cite{ref:Serafini17} started to attract significant attention~\cite{mele2025learning,Bittel_2025,fanizza2025,bittel2025energy,becker_cs,gandhari2023,conrad2024cvdesigns,oh2024entanglement,Liu_2025,coroi2025,fawzi2024optimalfidelity,Wu_2024,mobus2023dissipation,mobus2025heisenberg,Zhao_2025,mele2025symplecticrank,iosue2025,fanizza2025_cv_unitary,girardi2025gaussian,markovich2025,rosati2024learning,upreti2026exponentiallyimprovedeffectivedescriptionsphysical}.

Historically, the concept of tomography was conceived and first demonstrated experimentally in the continuous-variable setting in the early 1990s~\cite{smithey_measurement_1993,lvovsky2009continuous}, and only later became standard for finite-dimensional systems. In CV platforms, the basic measurement schemes are homodyne and heterodyne detection~\cite{ref:Serafini17}, from which one typically reconstructs phase-space representations such as the Wigner function, characteristic functions, or the Husimi function. While these techniques have been used experimentally for decades, they were long not accompanied by guarantees on the reconstruction error in trace distance, despite the trace distance having a clear operational interpretation as a notion of distinguishability between quantum states~\cite{ref:Helstrom76,ref:Holevo76Steklov}. With the advent of quantum learning theory, understanding the optimal performance---in particular, the sample complexity---of CV tomography has become an active research direction, which we review here.

\section{Tomography of non-Gaussian states with an energy constraint}
Continuous-variable systems are a key component of quantum science and technology, as they model bosonic and quantum-optical platforms~\cite{ref:Serafini17,ref:BV05,ref:WPGP12}. In the CV setting, the finite-dimensional notion of a \emph{qudit} is replaced by a \emph{mode}. The Hilbert space of a single mode is infinite-dimensional and can be written as $\mathcal{H}:=\mathrm{Span}\{|0\rangle,|1\rangle,\ldots,|d\rangle,\ldots\}$, where $|d\rangle$ denotes the Fock state with exactly $d$ photons~\cite{ref:Serafini17}. Equivalently, $\mathcal{H}$ can be identified with $L^2(\mathbb{R})$, the space of square-integrable complex-valued functions on $\mathbb{R}$. A CV system with $n$ modes is therefore described by the tensor-product Hilbert space $\mathcal{H}^{\otimes n}$.

The infinite dimensionality of $\mathcal{H}$ has an immediate consequence for learning: if the unknown state $\rho$ is allowed to be completely arbitrary on $\mathcal{H}^{\otimes n}$, then tomography has, strictly speaking, infinite sample complexity. Intuitively, without further assumptions the number of effective degrees of freedom is unbounded, so no finite number of copies can guarantee reconstruction with bounded error. In this sense, without prior information, tomography of CV quantum states is impossible.

In practice, however, CV experiments come with natural physical priors. The most common one is an \emph{energy constraint}, typically expressed via the mean photon number. Let $\hat{n}:=\sum_{k=0}^{\infty} k\,|k\rangle\!\langle k|$ be the single-mode number operator, and define the total photon-number operator on $n$ modes as $\hat{N}_n:=\sum_{i=1}^{n}\hat{n}^{(i)}$, where $\hat{n}^{(i)}$ acts as $\hat{n}$ on mode $i$ and as the identity on the remaining modes. We assume that the unknown state $\rho$ satisfies an energy-constraint promise of the form $\mathrm{Tr}[\rho\,\hat{N}_n]\le nE$, where $E$ upper bounds the mean photon number \emph{per mode}. This normalisation reflects the extensivity of energy and avoids an implicit dependence on the system size, since the constraint scales linearly with $n$.

In~\cite{mele2025learning} we determine the relevant scaling of the sample complexity of learning $n$-mode \emph{pure} states $\psi$ on $\mathcal{H}^{\otimes n}$ under the energy constraint $\mathrm{Tr}[\psi\,\hat{N}_n]\le nE$. Focusing on the dependence on the number of modes $n$ (and suppressing factors that do not scale explicitly with $n$), we obtain $N=\tilde{\Theta}\!\left(\frac{E^n}{\varepsilon^{2n}}\right)$. This shows that CV tomography is highly inefficient: the required number of copies grows exponentially with the system size $n$ (as it already does for finite-dimensional systems of $n$ qudits), but it also displays a dramatic dependence of $\varepsilon^{-2n}$ on the accuracy parameter $\varepsilon$. In particular, this $\varepsilon^{-2n}$ scaling is specific to CV tomography and stands in sharp contrast to the finite-dimensional $\varepsilon^{-2}$ scaling. Consequently, improving the error $\varepsilon$ by a constant factor in the CV setting entails an overhead that is exponential in the number of modes.

To illustrate the magnitude of this effect, consider tomography with target error $\varepsilon=10\%$ and assume that one copy of the state can be produced and processed every $1\,\mathrm{ns}$ (a typical timescale for qubits and optical pulses). Then the sample complexity formula for tomography of $10$-qubit pure states would lead to total tomography time of about $0.1\,\mathrm{ms}$. In contrast, the above result of~\cite{mele2025learning} implies that tomography of $10$-mode pure states with mean photon number per mode at most $1$ (so, a fair comparison with qubit systems) requires at least $\sim 3000\,\mathrm{years}$, showing that CV tomography becomes impractical already for a moderate number of modes.

Finally,~\cite{mele2025learning} also provides bounds for the sample complexity of learning mixed energy-constrained states, as well as an extension of the analysis to constraints on higher moments of the energy.

\section{Tomography of Gaussian states}
In the previous section we saw that tomography of arbitrary CV states is very inefficient, even in the presence of energy constraints. In many practical scenarios, however, one has the additional prior promise that the unknown CV state is Gaussian. This raises a natural question: can tomography of Gaussian states be performed efficiently, i.e., with sample complexity polynomial in the number of modes? In~\cite{mele2025learning} it was shown that this is indeed the case, and subsequent works~\cite{Bittel_2025,fanizza2025,ref:Holevo24TN,bittel2025energy,ref:CHMF+26,dimitroff2026learninggaussianopticalstates} further refined the picture.

By definition, a Gaussian state is a Gibbs state of a positive {quadratic} Hamiltonian in the position and momentum operators of an $n$-mode system; we refer to Ref.~\cite{ref:Serafini17} for a formal treatment. Gaussian states can be viewed as the quantum analogue of classical Gaussian distributions: they are uniquely determined by their \emph{first moments} and their \emph{covariance matrix}. Concretely, for an $n$-mode state $\rho$ one defines the quadrature vector $\hat{\mathbf{R}}:=(\hat{x}_1,\hat{p}_1,\ldots,\hat{x}_n,\hat{p}_n)^{\intercal}$, and the first moments and covariance matrix as $\mathbf{m}(\rho):=\mathrm{Tr}[\hat{\mathbf{R}}\,\rho]$ and $V(\rho):=\mathrm{Tr}\!\left[\{\hat{\mathbf{R}}-\mathbf{m}(\rho),(\hat{\mathbf{R}}-\mathbf{m}(\rho))^{\intercal}\}\rho\right]$, where $\{\cdot,\cdot\}$ denotes the anti-commutator. Importantly, these moments can be estimated experimentally via measurements of quadratures (homodyne detection), and for Gaussian states also via heterodyne detection~\cite{ref:Serafini17}.

At first sight this suggests a very simple learning strategy: estimate $\mathbf{m}(\rho)$ and $V(\rho)$ and reconstruct the corresponding Gaussian state. The subtle point is that we want guarantees in \emph{trace distance}, rather than in some metric on moments. Thus, one needs bounds that control the trace distance between two Gaussian states in terms of suitable distances between their first two moments. This is the starting point of the first works on Gaussian-state learning~\cite{mele2025learning,Bittel_2025,fanizza2025,ref:Holevo24TN,bittel2025energy}, and we review the best available bounds in the next subsection. Using this approach, the baseline protocol---estimate the first two moments via heterodyne detection and propagate the resulting error to trace distance (often called \emph{heterodyne tomography})---yields the following guarantee: restricting to heterodyne detection only, $N=\Theta(E^2 n^3/\varepsilon^2)$ copies are necessary~\cite{ref:CHMF+26} and sufficient~\cite{bittel2025energy} to perform tomography of an $n$-mode Gaussian state promised to have mean energy per mode at most $E$, within trace-distance error $\varepsilon$.

This bound is already polynomial in $n$ and $E$, and it is achieved by an operationally simple protocol (even though the trace-distance analysis is delicate). A natural next question is whether one can improve the energy dependence. In~\cite{bittel2025energy} we show that one can: it suffices to use $O\left(n^3/\varepsilon^2+(n+\log\log \log E)\log\log E\right)$ copies, which is essentially independent of $E$ up to double-logarithmic factors. The protocol relies on a simple idea: when the state is highly squeezed, heterodyne detection becomes inefficient because some quadrature variances are very small. The solution is to first learn the squeezing directions, then iteratively apply adaptive ``unsqueezing'' operations, and only at the end---once the state is approximately unsqueezed---apply heterodyne detection to estimate the moments. The protocol uses only Gaussian resources (adaptive passive Gaussian operations, squeezed states, and homodyne detection)~\cite{bittel2025energy}. In~\cite{ref:CHMF+26} we also prove that \emph{any} tomography algorithm restricted to Gaussian operations must use at least $\Omega(n^3/\varepsilon^2)$ copies. In other words, within the class of Gaussian operations the sample complexity of tomography of Gaussian states is $\tilde{\Theta}(n^3/\varepsilon^2)$, and this is optimal up to $\log\log E$ factors. Interestingly,~\cite{ref:CHMF+26} shows that if the Gaussian state is additionally promised to be pure, then the scaling improves to $\tilde{\Theta}(n^2/\varepsilon^2)$.

This naturally leads to the question of what happens beyond Gaussian operations. In~\cite{ref:CHMF+26} we prove a general lower bound of $\Omega(n^2/\varepsilon^2)$ that applies to \emph{any} tomography algorithm, including non-Gaussian ones. It is therefore natural to ask whether a non-Gaussian protocol can achieve this scaling and outperform all Gaussian strategies (which are constrained by the $\Omega(n^3/\varepsilon^2)$ lower bound~\cite{ref:CHMF+26}). While this question is still open for general Gaussian states, it can be resolved for squeezing-free Gaussian states, i.e., \emph{passive Gaussian states}~\cite{ref:CHMF+26}. In that setting, Gaussian operations still require $\Theta(n^3/\varepsilon^2)$ copies~\cite{ref:CHMF+26}, whereas a \emph{non-Gaussian} protocol achieves $\tilde{\Theta}(n^2/\varepsilon^2)$ copies~\cite{ref:CHMF+26}, giving a provable polynomial advantage of non-Gaussian algorithms over Gaussian ones. The key ingredient behind this non-Gaussian algorithm is the \emph{random purification channel}~\cite{ref:TWZ25,ref:PSTW25,ref:GML25}, a recently introduced tool originally developed in the finite-dimensional setting. Informally, it is a channel that takes $N$ i.i.d.\ copies of a mixed state and outputs a uniform convex combination of $N$ i.i.d.\ copies of one of its purifications. This idea was later adapted to the passive Gaussian setting~\cite{ref:WW25,ref:MGCF+25}, yielding a channel that transforms $N$ i.i.d.\ copies of a mixed passive Gaussian state into a uniform convex combination of $N$ i.i.d.\ copies of one of its \emph{Gaussian} purifications. Following the perspective of~\cite{ref:PSTW25}, where the random purification channel reduces mixed-state tomography to pure-state tomography in the qudit setting, in~\cite{ref:CHMF+26} we combine the passive random purification channel with pure Gaussian-state tomography and obtain a sample complexity $\tilde{\Theta}(n^2/\varepsilon^2)$ for learning passive Gaussian states. Since the random purification channel is inherently \emph{non-Gaussian}, this optimal scaling is achieved by a non-Gaussian protocol that provably outperforms all Gaussian approaches.

The task of learning an $n$-mode Gaussian state in trace distance can be seen as the quantum analogue of the classical problem of learning a $2n$-variate Gaussian probability distribution in total variation distance (TVD), which has sample complexity $\Theta(n^2/\varepsilon^2)$~\cite{ref:ABDHLMP20}. Interestingly, up to $\log\log E$ factors, the same scaling $\tilde{\Theta}(n^2/\varepsilon^2)$ also appears as the sample complexity for learning the Wigner function of a Gaussian state in TVD~\cite{ref:CHMF+26}, despite the fact that the Wigner function cannot be sampled from directly because of the Heisenberg uncertainty principle.

\section{Trade-off between learning efficiency and non-Gaussianity}
Above we have seen that, while tomography of arbitrary non-Gaussian CV states is extremely inefficient, tomography of Gaussian states can be performed efficiently. It is therefore natural to ask how the learning efficiency depends on the non-Gaussianity of the unknown state. In~\cite{mele2025learning} this trade-off is studied via tomography of \emph{$t$-doped Gaussian states}, i.e., pure $n$-mode states obtained from the vacuum by applying arbitrary Gaussian unitaries interleaved with at most $t$ single-mode non-Gaussian gates. The class reduces to Gaussian states when $t=0$, and it expands with $t$, approaching the full family of non-Gaussian states as $t\to\infty$. In this setting, the dependence of the sample complexity on $t$ directly captures how the efficiency of tomography deteriorates with increasing non-Gaussianity; Ref.~\cite{mele2025learning} shows that it grows exponentially in $t$, thereby interpolating between the efficient Gaussian regime and the extreme inefficiency of fully general non-Gaussian tomography discussed above.

This notion of non-Gaussianity (the parameter $t$ of a $t$-doped Gaussian state) was later generalised into a rigorous measure of non-Gaussianity in~\cite{mele2025symplecticrank}, the \emph{symplectic rank}. For a pure state, the symplectic rank is defined as the number of symplectic eigenvalues that are strictly larger than those of the vacuum. In~\cite{mele2025symplecticrank} it is proved that the sample complexity of tomography scales exponentially with the symplectic rank, making the dependence of tomography efficiency on non-Gaussianity explicit. The symplectic rank is a proper non-Gaussianity monotone, in the sense that it is non-increasing under post-selected Gaussian operations~\cite{mele2025symplecticrank}.

\section{Bosonic trace distance bounds}
A central technical ingredient in the analysis of quantum learning theory for bosonic systems is the use of inequalities that relate the trace distance between two states to suitable distances between their moments~\cite{mele2025learning,Bittel_2025,fanizza2025,ref:Holevo24TN,bittel2025energy,mele2025symplecticrank,ref:CHMF+26,ref:MBGF25,ref:MCQ26}.

Let $\rho(V,\mathbf{m})$ denote a Gaussian state with covariance matrix $V$ and first moment $\mathbf{m}$. After several refinements~\cite{mele2025learning,Bittel_2025,fanizza2025,ref:Holevo24TN,bittel2025energy}, the tightest known upper bound on the trace distance between two Gaussian states was obtained in~\cite{bittel2025energy} and reads $\|\rho(V,\mathbf{m})-\rho(W,\mathbf{t})\|_1 \le \frac{1+\sqrt{3}}{4}\,\mathrm{Tr}\!\left[(V^{-1}+W^{-1})\,|V-W|\right] + \|V^{-1/2}(\mathbf{m}-\mathbf{t})\|_2$. See also~\cite{ref:MBGF25,ref:MCQ26} for algorithms that compute the trace distance between two Gaussian states up to a prescribed precision. Moreover, a lower bound is given by $\|\rho(V,\mathbf{m})-\rho(W,\mathbf{t})\|_1 \ge \frac{1}{50}\min\!\left(1,\frac{\|V-W\|_2}{1+4\min(\|V\|_\infty,\|W\|_\infty)}\right)$~\cite{mele2025learning}, and additional lower bounds were introduced in~\cite{ref:CHMF+26}.

Ref.~\cite{ref:CHMF+26} also relates the trace distance between two $n$-mode Gaussian states $\rho,\sigma$ to the total variation distance between their Wigner functions $W_\rho,W_\sigma$. Specifically, it is shown that $\Omega(\mathrm{TVD}(W_\rho,W_\sigma)) \le \|\rho-\sigma\|_1 \le O\!\left(\sqrt{n}\,\mathrm{TVD}(W_\rho,W_\sigma)\right)$, and that the $\sqrt{n}$ factor in the upper bound can be removed when $\rho$ and $\sigma$ are pure.

Related trace-distance bounds in the non-Gaussian setting were derived in~\cite{mele2025symplecticrank}. For instance, the trace distance between two states $\rho,\sigma$ can be lower bounded in terms of their covariance matrices $V,W$ as $\|\rho-\sigma\|_1 \ge \frac{\|V-W\|_\infty^2}{1459\,\max\left(\mathrm{Tr}[\rho\,\hat{E}^2],\,\mathrm{Tr}[\sigma\,\hat{E}^2]\right)}$, where $\hat{E}:=\frac12\sum_{j=1}^n(\hat{x}_j^2+\hat{p}_j^2)$ is the energy operator.

\section{Testing whether an unknown state is Gaussian or not}
Given the prominent role of Gaussian states in quantum optics and CV-based quantum technologies, it is fundamental to understand how efficiently one can test whether an unknown CV state is Gaussian or not. Equivalently, one can ask for the minimum number of copies required to distinguish whether an unknown state is $\varepsilon_A$-close to the set of Gaussian states or $\varepsilon_B$-far from it (e.g., in trace distance). This question is studied in~\cite{girardi2025gaussian}. In the pure-state setting,~\cite{girardi2025gaussian} shows that a polynomial number of copies (polynomial in the number of modes and in the energy) suffices to decide whether the unknown state is close to the set of pure Gaussian states or far from it. In contrast, for general mixed states the same work proves that Gaussianity testing necessarily requires exponentially many copies, revealing a fundamental limitation for testing CV systems.

\section{Other topics in learning bosonic systems}
Beyond the topics discussed above, quantum learning theory for bosonic systems also encompasses tomography of structured families of non-Gaussian states~\cite{Zhao_2025,iosue2025,markovich2025,upreti2026exponentiallyimprovedeffectivedescriptionsphysical}, learning bosonic Gaussian unitaries~\cite{fanizza2025_cv_unitary}, bosonic variants of classical shadows~\cite{becker_cs,gandhari2023,conrad2024cvdesigns}, fidelity estimation from restricted measurement models~\cite{fawzi2024optimalfidelity}, provable quantum advantages in learning tasks~\cite{oh2024entanglement,Liu_2025,coroi2025,ref:CDK26}, and Hamiltonian learning in bosonic settings~\cite{fanizza2025,mobus2023dissipation,mobus2025heisenberg}.

\section{Open problems}
We conclude this review by listing selected open problems in quantum learning theory with bosonic systems. First, it is interesting to determine the ultimate sample complexity of tomography of energy-constrained states~\cite{mele2025learning}, Gaussian states~\cite{mele2025learning,Bittel_2025,fanizza2025,ref:Holevo24TN,bittel2025energy,ref:CHMF+26}, and Gaussianity testing~\cite{girardi2025gaussian}. Moreover, it would be interesting to understand whether any Gaussian protocol for tomography of Gaussian states is fundamentally required to exhibit an energy dependence---or whether the mild $\log\log E$ energy dependence of~\cite{bittel2025energy} can be removed: intuitively, the Heisenberg uncertainty principle seems to imply that this energy dependence is somehow necessary for Gaussian protocols. Moreover, it would be interesting to generalise the random purification channel to the general (non-passive) Gaussian case~\cite{ref:WW25,ref:MGCF+25}, which would have implications for finding the optimal sample complexity of tomography of Gaussian states~\cite{ref:CHMF+26}. Another important direction for future work concerns tomography of Gaussian processes: at present, (non-tight) upper bounds are known only in the restricted setting of Gaussian unitary channels~\cite{fanizza2025_cv_unitary}, while the general case of non-unitary Gaussian channels remains completely open. A promising way to address this is to generalise the construction of the random Stinespring superchannel to the Gaussian case~\cite{ref:GMZFL25,ref:YNM25}.

\end{document}